\documentstyle[12pt]{article}
\begin{document}

\title {ON THE DUAL INTERPRETATION OF
ZERO-CURVATURE FRIEDMANN-ROBERTSON-WALKER MODELS}
\author{L. Herrera\thanks{Also at Departamento de F\'\i sica,
U.C.V., Venezuela; email: lherrera@gugu.usal.es }\\
\'Area de F\'\i sica Te\'orica\\
Facultad de Ciencias\\
Universidad de Salamanca\\
37008, Salamanca, Spain.\\
\\
A. Di Prisco\thanks{On leave; email: lherrera@gugu.usal.es}\,\\
Departamento de F\'\i sica\\
Facultad de Ciencias\\ Universidad Central de Venezuela\\
Caracas, Venezuela.\\
\\
J. Ib\'a\~nez\thanks{email: wtpibmej@lg.ehu.es}\, \\
Departamento de F\'\i sica Te\'orica\\
Universidad del Pa\'\i s Vasco, Apdo.644,\\
48080 Bilbao, Spain.
}
\date{}
\maketitle

\newpage
\begin{abstract}
Two possible interpretations of FRW cosmologies (perfect fluid or
dissipative fluid)
are considered as consecutive phases of the system. Necessary conditions
are found,
for the transition from perfect fluid to dissipative regime to occur,
bringing out
the conspicuous role played by a particular state of the system (the
``critical point'').
\end{abstract}

\section{Introduction}
It is already a well stablished fact that a variety of line elements may
satisfy
the Einstein equations for different (physically meaningful) stress-energy
tensors (see \cite{Var} and references therein).

Particularly interesting is the situation with Friedmann-Robertson-Walker (FRW)
models, which, as it has been shown (\cite{CoTu},\cite{Co}, and references
therein),
do not necessarily represent perfect fluid solutions, but can also be exact
solutions for viscous and heat conducting fluids, with or without an
electromagnetic field.

Thus, for example, it has been shown \cite{CoTu}, that the zero-curvature
FRW metric
\begin{equation}
ds^2=-dt^2 + R^2(t) \left(dr^2 + r^2 d\theta^2 + r^2 \sin^2{\theta}
d\phi^2\right)
\label{el}
\end{equation}
satisfies Einstein equations, not only for perfect fluid matter distributions
\begin{equation}
\bar{T}_{\mu\nu}= (\bar\rho + \bar p) v_\mu v_\nu + \bar p g_{\mu\nu},
\label{Tpf}
\end{equation}
but also, for the stress-energy tensor of a magnetohydrodynamic viscous
fluid with heat conducion, viz.
\begin{equation}
{T}_{\mu\nu}={E}_{\mu\nu} + (\rho + p - \xi\Theta)u_\mu u_\nu + (p -
\xi\Theta) g_{\mu\nu} - 2\eta \sigma_{\mu\nu}
+ q_\mu u_\nu + q_\nu u_\mu
\label{Tc}
\end{equation}
where ${E}_{\mu\nu}, \rho, p, \xi, \eta, \Theta, \sigma_{\mu\nu}$ and
$q_\mu$ denote
respectively, the electromagnetic stress-energy tensor, the energy density,
the pressure,
the bulk and shear viscosity coefficients, the expansion, the shear tensor
and the
heat flux vector.

A fundamental difference between both interpretations of the models resides
in the fact that observers
moving along $v^{\mu}$ (perfect fluid case) are comoving with the fluid,
whereas those moving along
$u^\mu$ (dissipative case) detect a radial velocity of fluid particles
(tilted models) \cite{TrPa}. In other words, in the dissipative models the
source is necessarily non-comoving
or tilting relative to the geometrically ``preferred'' Ricci eigenvector
\cite{Co}. According to Coley and Tupper
\cite{CoTu} this tilting velocity  might be related to the observed motion
of our galaxy relative to the microwave
background radiation (CMB), and seems to be consistent with measurements of
the dipole anisotropy of the CMB temperature
distribution \cite{Co}.

A detailed analysis of many of those models shows that their observational
predictions
are in good agreement with available data \cite{Co}.

In this work, both interpretations are not considered as alternative,
mutually exclusive
possibilities, but rather as consecutive phases in the evolution of the
model. More specifically,
we shall consider the case when, initially, the system evolves as perfect
fluid without dissipation, and then at some instant starts to dissipate,
tending eventually again to a perfect fluid as $t \rightarrow \infty$.
The question we want to answer here is: what could be the necessary
condition for that transition to occur?.

For simplicity we shall consider neither electromagnetic field
($E_{\mu\nu}=0$) nor bulk  viscosity
($\xi=0$). Furthermore, the viscosity term will be modeled through an
anisotropic fluid and only transport equation for the
heat flux will be needed. The reason of this simplification becomes clear
later.Also, it should be kept in mind that the approach
used in this work only provides a "snapshot" of the system at the moment it
enters the dissipative regime, thereby giving
only a hint about its tendency, and not a complete description of its
evolution.All these simplifications represent the price to pay
in order to deal with mathematically tractable equations.

Two different initial situations will be considered:
\begin{enumerate}
\item The system initially evolves strictly as a perfect fluid. By this we
mean that there is not
dissipation whatsoever and preferred observers (associated with the four
velocity of the fluid)
are (strictly) comoving with the fluid.
\item The system initially evolves very close to the perfect fluid regime,
but there is still an small
(in a sense to be defined below) amount of dissipation, because of which,
preferred observers are
not strictly comoving with the fluid, but detect an small radial motion of
the fluid, so that quadratic and higher
powers, as well as time derivatives, of radial velocity are neglected
during this ``quasi-perfect fluid''regime.
\end{enumerate}

Then it is assumed that at some instant (say $t=\tilde t$), our system
starts to deviate from either of the initial
regimes and enters into dissipative regime.
In the first case the system leaves the perfect fluid condition by starting
to dissipate.
In the second case the system abandons the quasi-perfect fluid regime by
increasing the rate of dissipation.

We shall find conditions for transitions described above to occur. With
this purpose we shall evaluate
field equations and transport equations immediately after $\tilde t$, where
immediately means on a time
scale of the order of relaxation time. We shall see that the so called
``critical point'' plays a fundamental role in
the occurrence of the transition to the dissipative regime.

The critical point \cite{Hetal}--\cite{HeMaC}, corresponds to the situation
when (in relativistic units)

\begin{equation}
\frac{\kappa T}{\tau (\rho + p)}=1
\label{cp}
\end{equation}
where $\kappa, T$ and $\tau$ denote heat conduction coefficient,
temperature and relaxation time respectively.

It has been shown \cite{Hetal} that under that condition, the effective
inertial mass of any fluid element,
just after the system departs from equilibrium, vanishes. Also, it has been
shown that as a self-gravitating
system approaches the critical point, its active gravitational mass behaves
in such a way that enhances
the instability of the system \cite{HeDP}.

It will be shown below that transition from perfect fluid to dissipative
regime is allowed only if
the system is at the critical point. However, transition from the
quasi-perfect fluid phase
to a dissipative regime is, in principle, always possible, and closer is
the system to the critical point,
faster is the transition.

\section{Field and Transport equations}
\subsection{Field equations}
First of all, note that for any locally-Minkowskian observer, comoving with
the fluid, the four velocity
takes the obvious form
\begin{equation}
\tilde{u}^\mu = (1,0,0,0),
\label{ut}
\end{equation}
then, performing a Lorentz boost to the frame with respect to which a fluid
element has radial velocity
$\omega$, we find (in the coordinate system of (\ref{el}))
\begin{equation}
{u}^\mu = \left(\frac{1}{\sqrt{1-\omega^2}},\frac{\omega
R^{-1}}{\sqrt{1-\omega^2}},0,0\right)
\label{um}
\end{equation}
which is the same expression (2.1) of \cite{CoTu}, with their $\alpha$'s
and $\beta$'s defined in terms of
$\omega$ by
\begin{equation}
\alpha \equiv \frac{1}{\sqrt{1-\omega^2}},
\label{al}
\end{equation}
\begin{equation}
\beta \equiv \frac{\omega}{\sqrt{1-\omega^2}}.
\label{be}
\end{equation}
Next, the heat flux vector $q^\mu$, satisfying $q^\mu u_\mu = 0$, is given by
\begin{equation}
q_\mu = Q \left(\frac{\omega}{\sqrt{1-\omega^2}},
\frac{-R}{\sqrt{1-\omega^2}}, 0, 0\right)
\label{qm}
\end{equation}
which is the same expression (2.3) of \cite{CoTu}, with $Q^2 = q_\mu q^\mu$.

Then the energy-momentun tensor (\ref{Tc}) of the source may be writen as
\begin{equation}
T_{\mu\nu} = (\rho+P_\bot) u_\mu u_\nu + P_\bot g_{\mu\nu} + (P_r - P_\bot)
s_\mu s_\nu +
q_\mu u_\nu + u_\mu q_\nu
\label{Tmn}
\end{equation}
where $P_r$ and $P_\bot$ denote the two principal (anisotropic) stresses,
linked to $p$ and $\eta \sigma_{\mu\nu}$
by relations specified below, and the vector $s_\mu$ satisfies conditions
\begin{equation}
s^\mu s_\mu = 1 \qquad ; \qquad s^\mu u_\mu = 0 \qquad ; \qquad T_{\mu\nu}
s^\mu s^\nu = P_r
\label{sc}
\end{equation}

Then, Einstein equations read (in relativistic units and omitting the
$8\pi$ factor)

\begin{equation}
\frac{3 \dot{R}^2}{R^2} = \frac{1}{1-\omega^2} [\rho + P_r \omega^2 - 2 Q
\omega]
\label{E1}
\end{equation}
\begin{equation}
- \left(\frac{2 \ddot{R}}{R} + \frac{\dot{R}^2}{R^2}\right) =
\frac{1}{1-\omega^2} [\rho \omega^2 + P_r  - 2 Q \omega]
\label{E2}
\end{equation}
\begin{equation}
- \left(\frac{2 \ddot{R}}{R} + \frac{\dot{R}^2}{R^2}\right) = P_\bot
\label{E3}
\end{equation}
\begin{equation}
0 = (\rho + P_r) \omega - Q (1 + \omega^2)
\label{E4}
\end{equation}
where dots denote derivative with respect to $t$.

Comparing (\ref{Tc}) and (\ref{Tmn}) the following relations follow
\begin{equation}
P_r = p - \frac{2 \eta}{R^2} \sigma_{11} (1-\omega^2)
\label{rpr}
\end{equation}
\begin{equation}
P_\bot = p + \frac{2 \eta}{R^2} \sigma_{11} (1-\omega^2)
\label{rpt}
\end{equation}
\begin{equation}
\sigma_{00} = \frac{2}{3} \frac{\omega^2}{(1-\omega^2)} X = \frac{\omega^2
\sigma_{11}}{R^2} =
- \frac{2 \omega^2 \sigma_{22}}{r^2 R^2 (1-\omega^2)} = - \frac{\omega}{R}
\sigma_{01}
\label{rsig}
\end{equation}
where $X$ is given by \cite{CoTu}
\begin{equation}
X = \frac{\omega \dot\omega - \omega^2 \omega' R^{-1} - \omega R^{-1}
r^{-1} (1 - \omega^2)}{(1 - \omega^2)^{3/2}}
\label{X}
\end{equation}
and prime denotes derivative with respect to $r$.

Observe that from (\ref{E2}) and (\ref{E3}), one obtains
\begin{equation}
(\rho + P_\bot) \omega^2 + (P_r - P_\bot) - 2Q\omega = 0
\label{tc}
\end{equation}
and using (\ref{E4}) and (\ref{tc}) it follows
\begin{equation}
- \frac{(P_r - P_\bot)}{\omega} + Q (1 - \omega^2) = 0
\label{qv}
\end{equation}

Thus, if $P_r = P_\bot$ (which implies by virtue of
(\ref{rpr})--(\ref{rsig}), $\eta \sigma_{\mu\nu} = 0$),
then $\omega = 1$, which is obviously inadmissible. Therefore, pure heat
conduction (plus a fluid) is not
consistent with FRW metric.

After a simple algebra, field equations (\ref{E1})--(\ref{E4}) may be
written as
\begin{equation}
\rho = \frac{3 \dot{R}^2}{R^2} + (P_r - P_\bot)
\label{ro}
\end{equation}
\begin{equation}
P_\bot = - \left(\frac{2 \ddot{R}}{R} + \frac{\dot{R}^2}{R^2}\right)
\label{Pt}
\end{equation}
\begin{equation}
P_r = - \left(\frac{2 \ddot{R}}{R} + \frac{\dot{R}^2}{R^2}\right) + Q \omega
\label{Pr}
\end{equation}
\begin{equation}
Q\left(1-\omega^2\right) = 2 \omega \left(\frac{\dot{R}^2}{R^2} -
\frac{\ddot{R}}{R}\right)
\label{Q}
\end{equation}
It is worth noticing that (\ref{Pt})--(\ref{Q}) imply that dissipation
takes place
only if $\omega \not= 0$ and $P_r \not= P_\bot$.

Later it will be convenient to write $R(t)$ in the form
\begin{equation}
R(t)=t^n
\label{Rt}
\end{equation}
where in general $n=n(t)$. Thus, $n=2/3$ reproduces the Einstein-de Sitter
model, whereas
\begin{equation}
n=\frac{H_0 t}{\ln{t}}
\label{nex}
\end{equation}
reproduces the exponential inflationary model with $R\sim e^{H_0 t}$,
$\rho + p = 0$ and $p = P_r = P_\bot$.

Using (\ref{Rt}), field equations (\ref{ro})--(\ref{Q}) become
\begin{equation}
\rho = 3 \left(\dot{n} \ln{t}\right)^2 + 6 n \dot{n} \frac{\ln{t}}{t} +
\frac{3 n^2}{t^2} + (P_r - P_\bot)
\label{ron}
\end{equation}
\begin{equation}
P_\bot = - 3 \left(\dot{n} \ln{t}\right)^2 - 6 n \dot{n} \frac{\ln{t}}{t} -
\frac{3 n^2}{t^2}
- 2 \ddot{n} \ln{t} - \frac{4 \dot{n}}{t} + \frac{2 n }{t^2}
\label{ptn}
\end{equation}
\begin{equation}
P_r = Q \omega - 3 \left(\dot{n} \ln{t}\right)^2 - 6 n \dot{n}
\frac{\ln{t}}{t} -
\frac{3 n^2}{t^2}
- 2 \ddot{n} \ln{t} - \frac{4 \dot{n}}{t} + \frac{2 n }{t^2}
\label{prn}
\end{equation}
\begin{equation}
Q\left(1-\omega^2\right) = \omega \left(\frac{2 n}{t^2} - \frac{4
\dot{n}}{t} - 2 \ddot{n} \ln{t}\right)
\label{Qn}
\end{equation}

\subsection{Transport equations}
In order to avoid the important drawbacks of Eckart's theory, we shall use
transport equations derived from the Israel-Stewart approach
\cite{Is}--\cite{Ma}.
However, since these equations are going to be evaluated immediately after the
system leaves the perfect fluid (or the quasi-perfect fluid) regime, then
the truncated version of the
theory leads to the same result as the full theory, and therefore the
former will be used here. For the same reason, the shear viscosity
contributions
result in terms which are negligible in the approximation used here (see
below).
Therefore we shall deal only with the transport equation for the heat flux,
which reads \cite{Ma}
\begin{equation}
q^\mu + \tau h^{\mu\beta} u^\gamma q_{\beta;\gamma} =
- \kappa \left(h^{\mu\beta} T_{,\beta} + T a^\mu\right)
\label{tr}
\end{equation}
where as usual, $h^{\mu\beta}$ denotes the projector onto the three space
orthogonal to $u^\mu$,
and $a^\mu$ denotes the four acceleration.

In terms of $n$, the only non vanishing independent component of (\ref{tr})
is the $r$-component,
which reads
\begin{eqnarray}
&-& \frac{Q}{t^n \left(1-\omega^2\right)^{1/2}} +
\tau \left\{- \frac{\dot Q}{t^n \left(1-\omega^2\right)} - \frac{Q'
\omega}{t^{2n} \left(1-\omega^2\right)}\right\}=
\nonumber \\
&-& \kappa \{\frac{\omega \dot T}{t^n \left(1-\omega^2\right)} +
\frac{T'}{t^{2n} \left(1-\omega^2\right)}
\nonumber \\
&+& T
\left[\frac{\dot \omega}{t^n \left(1-\omega^2\right)^{2}} +
\frac{\omega \omega'}{t^{2n} \left(1-\omega^2\right)^2} +
\frac{\omega}{t^n \left(1-\omega^2\right)} \left(\dot n \ln t +
\frac{n}{t}\right)\right] \}
\label{rco}
\end{eqnarray}
where prime denotes derivative with respect to $r$ (the $t$-component of
(\ref{tr}) is just (\ref{rco})
multiplied by $\omega$).

\section{Entering into dissipative regime}
As was already mentioned, an essential feature of the dissipative regime, as
implied by
(\ref{Q}) or (\ref{Qn}), is the non comoving nature of observers of the
congruence
$u^\mu$ ($\omega \not = 0$). This means that $\omega$ may be used as
control variable
to assess how far (or close) is the system from the perfect fluid regime.

So, let us now assume that for $t<\tilde t$, our system is in either the
perfect fluid regime
or in the quasi-perfect fluid regime. Both of which are characterized as
follows:
\begin{enumerate}
\item Perfect fluid regime
\begin{equation}
Q = \omega = 0
\label{pf}
\end{equation}
Preferred observers are strictly comoving with the fluid.
\item Quasi-perfect fluid regime
\begin{equation}
Q \approx \omega \approx O(\epsilon)
\label{qp1}
\end{equation}
\begin{equation}
\dot Q \approx \dot \omega \approx \omega^2 \approx Q \omega \approx Q^2
\approx 0
\label{qp2}
\end{equation}
Preferred observers are not strictly comoving with the fluid, but
(\ref{qp1}) and (\ref{qp2}) hold,
i.e. within this regime the system always remains ``close'' to the perfect
fluid phase. In both cases, of course, $r$ derivatives of $Q$ and $\omega$,
are of the same order of magnitude as these quantities.
\end{enumerate}

Next, let us assume that at $t = \tilde t$, the system is allowed to abandon
either of regimes (1. or 2.) and enter into a dissipative phase.
We shall now find the conditions for these transitions to occur.

Let us start with the case 1. Then, evaluating the transport equation
immediately
after
the system leaves the perfect fluid regime ($t \approx \tilde t + O(\tau)$),
we obtain that eq.(\ref{rco}) yields
\begin{equation}
\dot Q = \frac{\kappa T}{\tau} \dot \omega
\label{Qp}
\end{equation}
where (\ref{pf}) and the fact that $T' \approx O(\omega) \approx O(Q)$,
have been used.
Observe that immediately after the system leaves the perfect fluid regime 
\begin{equation}
P_r - P_\bot = Q \omega = O(\omega^2)
\label{PQ}
\end{equation}
therefore the contribution of the shear viscosity at $t=\tilde t$ is one
order smaller
than the pure heat conduction effects, and accordingly will not be
considered here.

Next, it will be more convenient to use the ``conservation laws'' (Bianchi
identities)
\begin{equation}
T^{\mu\nu}_{;\nu} = 0
\label{cl}
\end{equation}
instead of field equations.

The $r$-component of (\ref{cl}) reads
\begin{eqnarray}
T^{r \nu}_{;\nu} = 0 &=&
\frac{\left(\dot \rho + \dot P_r\right) \omega}{t^n \left(1 - \omega^2\right)}
- \frac{\dot Q \left(1 + \omega^2\right)}{t^n \left(1 - \omega^2\right)}
+ \frac{\left(\rho + P_r\right) \dot \omega \left(1 + \omega^2\right)}{t^n
\left(1 - \omega^2\right)^2}
- \frac{Q \, 4 \omega \dot \omega}{t^n \left(1 - \omega^2\right)^2}
\nonumber \\
&+& \frac{4}{t^n} \left(\dot n \ln{t} + \frac{n}{t}\right)
\left[\frac{\left(\rho + P_r\right) \omega}{1 - \omega^2} - \frac{Q \left(1 +
\omega^2\right)}{1 - \omega^2}\right]
\nonumber \\
&+& \frac{1}{t^{2n}} \left[\frac{\left(\rho + P_r\right) 2 \omega
\omega'}{\left(1 - \omega^2\right)^2}
- \frac{2 Q' \omega}{1 - \omega^2}-\frac{\left(1+\omega^2\right) 2 Q
\omega'}{\left(1 - \omega^2\right)^2}
\right]
\nonumber \\
&+& \frac{2}{r t^{2n}} \left[\frac{\left(\rho + P_\bot\right) \omega^2}{1 -
\omega^2} -
\frac{Q \omega}{1 - \omega^2}\right]
\label{rcl}
\end{eqnarray}
which evaluated immediately after $\tilde t$, yields
\begin{equation}
\dot Q = \left(\rho + P_r\right) \dot \omega
\label{dQd}
\end{equation}
Combining (\ref{Qp}) and (\ref{dQd}), we finally obtain
\begin{equation}
\dot \omega \left(\rho + P_r\right) \left(1 - \frac{\kappa T}{\tau \left(\rho
+ P_r\right)}\right) = 0.
\label{f}
\end{equation}
Observe that in (\ref{dQd}) and (\ref{f}) we may replace $P_r$ by $p$, since
they differ by terms of order $O(\omega^2)$.

Now, since initially ($t<\tilde t$) the system was evolving as a perfect
fluid ($\omega = Q =0$),
then a necessary condition for leaving that regime at $t = \tilde t$ is
\begin{equation}
\frac{\kappa T}{\tau \left(\rho + p\right)} = 1
\label{nc}
\end{equation}
for otherwise $\dot \omega = 0$, and, as can be easily checked by taking
consecutive time derivatives of (\ref{rco}) and (\ref{rcl}), all higher
time derivatives of $\omega$ (and $Q$) will also vanish. As mentioned before,
eq.(\ref{nc}) defines the critical point.

Let us now consider the case 2. In this case the system (initially) is
evolving as a quasi perfect fluid, leaving that regime at $t = \tilde t$.
Evaluating (\ref{rco}) at $t<\tilde t$, we obtain (observe that with the
choice of (\ref{qm}),
$Q > 0$ implies that $q^r$ points inward)
\begin{equation}
Q = \kappa \left[\frac{T'}{t^n} + T \omega \left(\dot n \ln{t} +
\frac{n}{t}\right) + \omega \dot T\right]
\label{Q2}
\end{equation}
then evaluating the same equation (\ref{rco}) immediately after $\tilde t$,
and using (\ref{Q2}) we get
\begin{equation}
\dot Q = \frac{\kappa T}{\tau} \dot \omega
\label{Qp2}
\end{equation}
as in the previous case (eq.(\ref{Qp})).

Next, evaluating (\ref{rcl}) just after $\tilde t$ one obtains
\begin{equation}
\frac{\omega \left(\dot \rho + \dot P_r\right)}{t^n}
- \frac{\dot Q}{t^n}
+ \frac{\dot \omega \left(\rho + P_r\right)}{t^n}
+ \frac{4}{t^n} \left(\dot n \ln{t} + \frac{n}{t}\right) \left[\left(\rho +
P_r\right) \omega - Q\right]
= 0
\label{eq}
\end{equation}
or, using (\ref{ron})--(\ref{Qn}) and (\ref{Qp2})
\begin{equation}
\dot \omega \left(\rho + P_r\right) \left[1 - \frac{\kappa T}{\tau \left(\rho
+ P_r\right)}\right] =
- \omega \left(\dot \rho + \dot P_r\right)
\label{pen}
\end{equation}
In the case $n = $constant, (\ref{pen}) becomes (using (\ref{ron}) and
(\ref{prn}) and replacing $P_r$ by $p$)
\begin{equation}
\dot \omega = \frac{2 \omega }{\left(1 - \frac{\kappa T}{\tau \left(\rho +
p\right)}\right) t}
\label{ul}
\end{equation}
Therefore $\dot\omega \not = 0$, and is larger, the closer is the system to the
critical point.

\section{Conclusions}
We have assumed in this paper that the zero-curvature FRW metric represents
consecutively, first a perfect
(or quasi-perfect) fluid solution, and then a dissipative
solution of Einstein equations.

Necessary conditions for the transition from one state to another have been
found.

If the system is initially evolving as a pure perfect fluid, then
transition to the
dissipative regime demands that the fluid satisfies the critical point
condition.
One important question related to this case is what could be the physical
reasons for
the system to abandon the equilibrium state?. One possible scenario might
be the appearance
of a dissipative process like particle creation, which as is known, is
formally equivalent
to the introduction of viscous terms \cite{particle}. Another possibility
might be the decreasing of opacity
of the fluid from very high values,
preventing the propagation of null mass particles as photons and neutrinos
(trapping),
to smaller values allowing for radiative heat conduction and viscosity.A
similar situation occurs during
the gravitational collapse of stars (see \cite{Arnett} and references
therein) and the Kelvin-Helmholtz phase
of neutron stars formation (see \cite{Burrows} and  references therein).

Of course the full implementation of any of these, or any other, scenarios
would require a
much more elaborated setup than the one presented here.
The use of FRW with a single dissipative fluid is obviously very limiting,
however as we have seen,
it leads to mathematically tractable equations.

In the same order of ideas, observe that if the system is, in the initial
phase, inflating exponentially
with $\rho + P_r = 0$,
then as it follows from (\ref{dQd}) and (\ref{Qp2}), it will not leave that
regime as long as  $\rho + P_r = 0$.
This is an obvious consequence from (\ref{ron})--(\ref{Qn}) which implies
$Q = 0$ if $\rho + P_r = 0$, parenthetically it is worth noticing that the
exit from inflation is usually
explained in terms of dissipative process like particle production.
From the above, it follows that though the obtained results are independent
on the equation of state,
some specific cases are not allowed since they exclude
the possibility of dissipation (e.g. $\rho + P_r = 0$).

If the system is allowed to evolve initially as a quasi-perfect fluid, then
transition
to dissipative regime may always occur, without requiring the critical
point to be attained,
however such transition will be ``faster'', the closer is the system to the
critical point.

Another important question to ask is if real physical systems may reach the
critical point.
Indeed, causality and stability requirements obtained from a linear
perturbative scheme
are violated, close to, but below the critical point (in the absence of
viscosity) \cite{HeMa}.
On the other hand however, examples of fluids attaining the critical point
and exhibiting
reasonable physical properties, have been presented elsewhere
\cite{HeMaC},\cite{HeMaA}.

The possible solution to this apparent contradiction being the non-validity
of the linear
scheme used to obtain causality and stability conditions, close to the
critical point
(see a discussion on this point in \cite{HeMa}).
At any rate, in presence of both, heat conduction and viscosity, the
corresponding
critical point is beyond causality and stability conditions \cite{HeMaC}.

\section{Acknowledgments}
The authors are indebted to an anonymous referee for detecting an important
flaw in an earlier version of
this manuscript. This work was partially supported by the
Spanish Ministry of Education under grant PB96-0250,and PB96-1306.


\end{document}